\documentclass[10pt,journal,compsoc]{IEEEtran}
\usepackage{cite}
\usepackage{multirow}
\usepackage{graphicx}
\usepackage{color}
\usepackage{dcolumn}
\usepackage{amsmath}
\usepackage{kantlipsum}
\usepackage{url}
\usepackage{microtype}
\usepackage{verbatim}
\usepackage{colortbl}
\usepackage{arydshln}
\usepackage{multirow}
\newcolumntype{K}[1]{>{\centering\arraybackslash}p{#1}}
\usepackage{booktabs}
\usepackage[flushleft]{threeparttable}
\usepackage{hhline}
\usepackage{algorithm}
\usepackage{algorithmic}
\usepackage{mathtools}

\ifCLASSOPTIONcompsoc
\usepackage[caption=false, font=normalsize, labelfont=sf, textfont=sf]{subfig}
\else
\usepackage[caption=false, font=footnotesize]{subfig}
\fi
\begin{document}
\title{CovidDeep: SARS-CoV-2/COVID-19 Test Based on Wearable Medical Sensors and Efficient Neural 
Networks}

\author{Shayan~Hassantabar,~Novati~Stefano,~Vishweshwar~Ghanakota,~Alessandra~Ferrari,~Gregory~N.~Nicola, Raffaele~Bruno,~Ignazio~R.~Marino,~Kenza~Hamidouche,~and~Niraj~K.~Jha,~\IEEEmembership{Fellow,~IEEE}
\thanks{This work was supported by NSF under Grant No. CNS-1907381.
Shayan Hassantabar and Niraj K. Jha are with the Department of Electrical Engineering, and Kenza Hamidouche is with the Department of Operations Research and Financial Engineering, Princeton 
University, Princeton, NJ, 08544 USA, e-mail:\{seyedh,jha,kenzah\}@princeton.edu.
Novati Stefano and Alessandra Ferrari are with the Division of Infectious Diseases Unit, Fondazione
IRCCS Policlinico San Matteo, Pavia, Italy, e-mail: \{S.Novati,alessandra.ferrari\}@smatteo.pv.it.
Raffaele Bruno is with the Division of Infectious Diseases Unit, Fondazione IRCCS Policlinico San
Matteo, Pavia, Italy, and the Department of Medical, Surgical, Diagnostic and Pediatric Science, 
University of Pavia, Italy, e-mail: \{raffaele.bruno@unipv.it\}.
Vishweshwar Vinnakota and Gregory N. Nicola are with NeuTigers, Inc., Brooklyn, 
NY 11201, USA, e-mail: \{vishu,greg\}@neutigers.com. Ignazio R. Marino is with Thomas Jefferson 
University and Jefferson Health, Philadelphia, PA 19107, USA, e-mail: \{Ignazio.Marino@jefferson.edu\}.
}
}
\IEEEtitleabstractindextext{%
\begin{abstract}
The novel coronavirus (SARS-CoV-2) has led to a pandemic. Because of its highly contagious 
nature, it has spread rapidly, resulting in major disruption to public health and a huge loss
of human life.  In addition, due to governmental orders for isolation and social distancing, it has 
also had a severe negative impact on the world economy.  As a result, it is widely recognized that 
widespread testing is key to containing the spread of the disease and opening up the economy.  
However, the current testing regime based on Reverse Transcription-Polymerase Chain Reaction for 
SARS-CoV-2 has been unable to keep up with testing demands and also suffers from a relatively 
low positive detection rate in the early stages of the resultant disease, called COVID-19. Hence, 
there is a need for an alternative approach for repeated large-scale testing of SARS-CoV-2/COVID-19. 
The emergence of wearable medical sensors (WMSs) and deep neural networks (DNNs) points to a 
promising approach to address this challenge.  WMSs enable continuous and user-transparent monitoring 
of physiological signals. However, disease detection based on WMSs/DNNs and their deployment on 
resource-constrained edge devices remain challenging problems. To address these problems, we propose 
a framework called CovidDeep that combines efficient DNNs with commercially available WMSs for 
pervasive testing of the virus in both the asymptomatic and symptomatic cases.  CovidDeep does not 
depend on manual feature extraction. It directly operates on WMS data and some easy-to-answer 
questions in a questionnaire whose answers can be obtained through a smartphone application.  
We collected data from 87 individuals, spanning three cohorts that include healthy, asymptomatic 
(but SARS-CoV-2-positive) as well as symptomatic COVID-19 patients.  We trained DNNs on various 
subsets of the features automatically extracted from six WMS and questionnaire categories to perform 
ablation studies to determine which subsets are most efficacious in terms of test accuracy for a 
three-way classification.  The highest test accuracy we obtained was 98.1\%.  Since data collection 
was limited to only 87 individuals (because of the intensive nature of data collection), we also 
experimented with augmenting the real training dataset with a synthetic training dataset drawn from 
the same probability distribution.  We used the synthetic dataset to impose a prior on the DNN 
weights. Furthermore, we leveraged a grow-and-prune DNN synthesis paradigm to simultaneously learn 
both the weights and the network architecture. Addition of synthetic data and use of grow-and-prune 
synthesis boosted the accuracy of the various DNNs further and simultaneously reduced their size and 
floating-point operations. This makes the CovidDeep DNNs both accurate and efficient, in terms of 
memory requirements and computations.  The resultant DNNs can be easily deployed on edge devices, 
e.g., smartwatch or smartphone, which has the added benefit of preserving patient privacy.
\end{abstract}

\begin{IEEEkeywords}
COVID-$19$ test; grow-and-prune synthesis; neural networks; SARS-CoV-2; synthetic data generation; 
wearable medical sensors.
\end{IEEEkeywords}}
\maketitle

\IEEEdisplaynontitleabstractindextext

\IEEEpeerreviewmaketitle

\IEEEraisesectionheading{\section{Introduction}\label{sect:introduction}}

\IEEEPARstart {S}{ARS-CoV-2}, also known as novel coronavirus, emerged in China and soon after 
spread across the globe.  The World Health Organization (WHO) named the resultant disease 
COVID-$19$. COVID-19 was declared a pandemic on March 11, $2020$ 
\cite{world2020coronavirus}.  In its early stages, the symptoms of COVID-$19$ include fever, cough, 
fatigue, and myalgia.  However, in more serious cases, it can lead to shortness of breath, pneumonia, 
severe acute respiratory disorder, and heart problems, and may lead to death 
\cite{mahase2020coronavirus}.  It is of paramount importance to detect which individuals are 
infected at as early a stage as possible in order to limit the spread of disease through
quarantine and contact tracing.  In response to COVID-19, governments around the 
world have issued social distancing and self-isolation orders.  This has led to a significant increase 
in unemployment across diverse economic sectors. As a result, COVID-$19$ has triggered an 
economic recession in a large number of countries \cite{nicola2020socio}. 

Reverse Transcription-Polymerase Chain Reaction (RT-PCR) is currently the gold standard 
for SARS-CoV-2 detection \cite{butt2020deep}. This test is based on viral nucleic acid detection 
in sputum or nasopharyngeal swab.  Although it has high specificity, it has several drawbacks. 
The RT-PCR test is invasive and uncomfortable, and non-reusable testing kits have led to
significant supply chain deficiencies.  
SARS-CoV-2 infection can also be assessed with an antibody test \cite{dheda2020diagnosis}. 
However, antibody titers are only detectable from the second week of illness onwards and persist for an
uncertain length of time. The antibody test is also invasive, requiring venipuncture which, in
combination with a several-day processing time, makes it less ideal for rapid mass screening.
In the current economic and social situation, there is a great need for an alternative 
SARS-CoV-2/COVID-19 detection method that is easily accessible to the public for repeated testing
with high accuracy.

To address the above issues, researchers have begun to explore the use of artificial intelligence 
(AI) algorithms to detect COVID-$19$ \cite{bullock2020mapping}.  
Initial work concentrated on CT scans and X-ray images \cite{farooq2020covid, wang2020covid,
SIRM,giovagnonidiagnosi,butt2020deep,zhang2020covid,narin2020automatic,abbas2020classification,
hall2020finding,sethy2020detection,li2020artificial,gozes2020rapid,apostolopoulos2020covid,
wang2020fully,afshar2020covid, hassantabar2020diagnosis}. A survey of such datasets can be found in 
\cite{kalkreuth2020covid,cohen2020covid}.
These methods often rely on transfer learning of a convolutional neural network (CNN) architecture, 
pre-trained on large image datasets, on a smaller COVID-$19$ image dataset.  
However, such an image-based AI approach faces several challenges that include lack of large
datasets and inapplicability outside the clinic or hospital. In addition, other work
\cite{Lin2020} shows that it is difficult to distinguish COVID-19 pneumonia from influenza
virus pneumonia in a clinical setting using CT scans.  Thus, the work in this area is not
mature yet.

CORD-19 \cite{cord19} is an assembly of $59000$ scholarly articles on COVID-$19$.
It can be used with natural language processing methods to distill useful information on
COVID-$19$-related topics.

AI$4$COVID-$19$ \cite{imran2020ai4covid} performs a preliminary diagnosis of COVID-$19$ through
cough sample recordings with a smartphone application. However, since coughing is a common symptom of
two dozen non-COVID-$19$ medical conditions, this is an extremely difficult task. Nonetheless,
AI$4$COVID-$19$ shows promising results and opens the door for COVID-$19$ diagnosis through
a smartphone.

The emergence of wearable medical sensors (WMSs) offers a promising way to tackle these challenges. 
WMSs can continuously sense physiological signals throughout the day \cite{yin2017health}. 
Hence, they enable constant monitoring of the user's health status.  Training AI algorithms
with data produced by WMSs can enable pervasive health condition tracking and disease onset 
detection \cite{yin2019diabdeep}.  This approach exploits the knowledge distillation 
capability of machine learning algorithms to directly extract information from physiological signals. 
Thus, it is not limited to disease detection in the clinical scenarios. 

We propose a framework called CovidDeep for daily detection of SARS-CoV-2/COVID-19 based on
off-the-shelf WMSs and compact deep neural networks (DNNs).  It bypasses manual feature engineering 
and directly distills information from the raw signals captured by available WMSs.
It addresses the problem posed by small COVID-19 datasets by relying on intelligent synthetic data 
generation from the same probability distribution as the training data \cite{hassantabar2020Tutor}. 
These synthetic data are used to pre-train the DNN architecture in order to impose a prior on the 
network weights.  To cut down on the computation and storage costs of the model without any
loss in accuracy, CovidDeep leverages the grow-and-prune DNN synthesis paradigm 
\cite{dai2017nest, hassantabar2019scann}.  This not only improves accuracy, but also shrinks 
model size and reduces the computation costs of the inference process. 

The major contributions of this article are as follows:

\begin{itemize}
    \item We propose CovidDeep, an easy-to-use, accurate, and pervasive SARS-CoV-2/COVID-19 
detection framework. It combines features extracted from physiological signals using WMSs and 
simple-to-answer questions in a smartphone application-based questionnaire with efficient DNNs.
    \item It uses an intelligent synthetic data generation module to obtain a synthetic 
dataset \cite{hassantabar2020Tutor}, labeled by decision rules. The synthetic dataset is used to 
pre-train the weights of the DNN architecture.
    \item It uses a grow-and-prune DNN synthesis paradigm that learns both an efficient
architecture and weights of the DNN at the same time \cite{dai2017nest, hassantabar2019scann}. 
    \item It provides a solution to the daily SARS-CoV-2/COVID-19 detection problem. It captures all 
the required physiological signals non-invasively through comfortably-worn WMSs that are commercially 
available. 
\end{itemize}

The rest of the article is organized as follows. Section \ref{sect:related} reviews background
material.  Section \ref{sect:methodology} describes the CovidDeep framework. Section 
\ref{sect:implementation} provides implementation details. Section \ref{sect:results} presents 
experimental results.  Section \ref{discussion} provides a short discussion on CovidDeep and possible 
directions for future research. Finally, Section \ref{conclusion} concludes the article. 

\section{Background}
\label{sect:related}
In this section, we discuss background material related to the CovidDeep framework. 
It involves recent methods for synthesizing and training efficient DNN architectures.

One approach is based on the use of efficient building blocks. Using such blocks results in compact 
networks and significantly reduces the computational costs and storage needs.
For example, inverted residual blocks used in MobileNetV$2$ 
\cite{sandler2018mobilenetv2} reduce the number of parameters and the
floating-point operations (FLOPs) greatly. 
In addition, spatial convolution is one 
of the most computationally expensive operations in CNN architectures. 
To address this issue, ShuffleNet-v$2$ \cite{ma2018shufflenet} uses the depth-wise separable 
convolutions and channel-shuffling operations. Furthermore, Shift \cite{wu2018shift} addresses this problem by using shift-based modules that combine shifts and point-wise convolutions.
Neural architecture search (NAS) is also used in the literature to automatically generate compact architectures.
For example, FBNetV$2$ \cite{wan2020fbnetv2} uses differentiable NAS approach to synthesize compact CNN architectures. 
Efficient performance predictors, e.g., for accuracy, latency, and energy, are also used to accelerate the DNN search process \cite{dai2018chamnet, hassantabar2019steerage}. 
FBNetV$3$ \cite{dai2020fbnetv3} takes into account the training recipe (i.e., training hyperparameters) in the NAS as well, leading to finding higher accuracy-recipe combinations. 

In addition, DNN compression methods can remove redundancy in the DNN models. Network 
pruning~\cite{han2015deep} uses a pruning methodology to remove redundancy from both CNN and 
multilayer-perceptron architectures.  ESE \cite{han2017ese} shows the pruning methods are also 
helpful in removing redundancy in recurrent neural networks. Dai et al.~\cite{dai2017nest, 
dai2018grow} combine network growth with pruning to generate efficient CNNs and long short-term 
memories.  SCANN \cite{hassantabar2019scann} combines feature dimensionality reduction with 
grow-and-prune synthesis to generate very compact models that can be easily deployed on edge devices 
and Internet-of-Things sensors.

Orthogonal to the above works, low-bit quantization of DNN weights can also be used to reduce computations in a network with little to no accuracy drop \cite{zhu2016trained}.

\section{Methodology}
\label{sect:methodology}
In this section, we present the CovidDeep framework. First, we give an
overview of the entire framework. Then, we describe the DNN architecture that
is used in CovidDeep for inference. We also describe how synthetic data generation can be used
to impose a prior on the DNN weights and then use the DNN grow-and-prune synthesis paradigm to
boost the test accuracy further and ensure computational efficiency of the model. 

\begin{figure*}[!ht]
    \centering
    \includegraphics[scale=0.5]{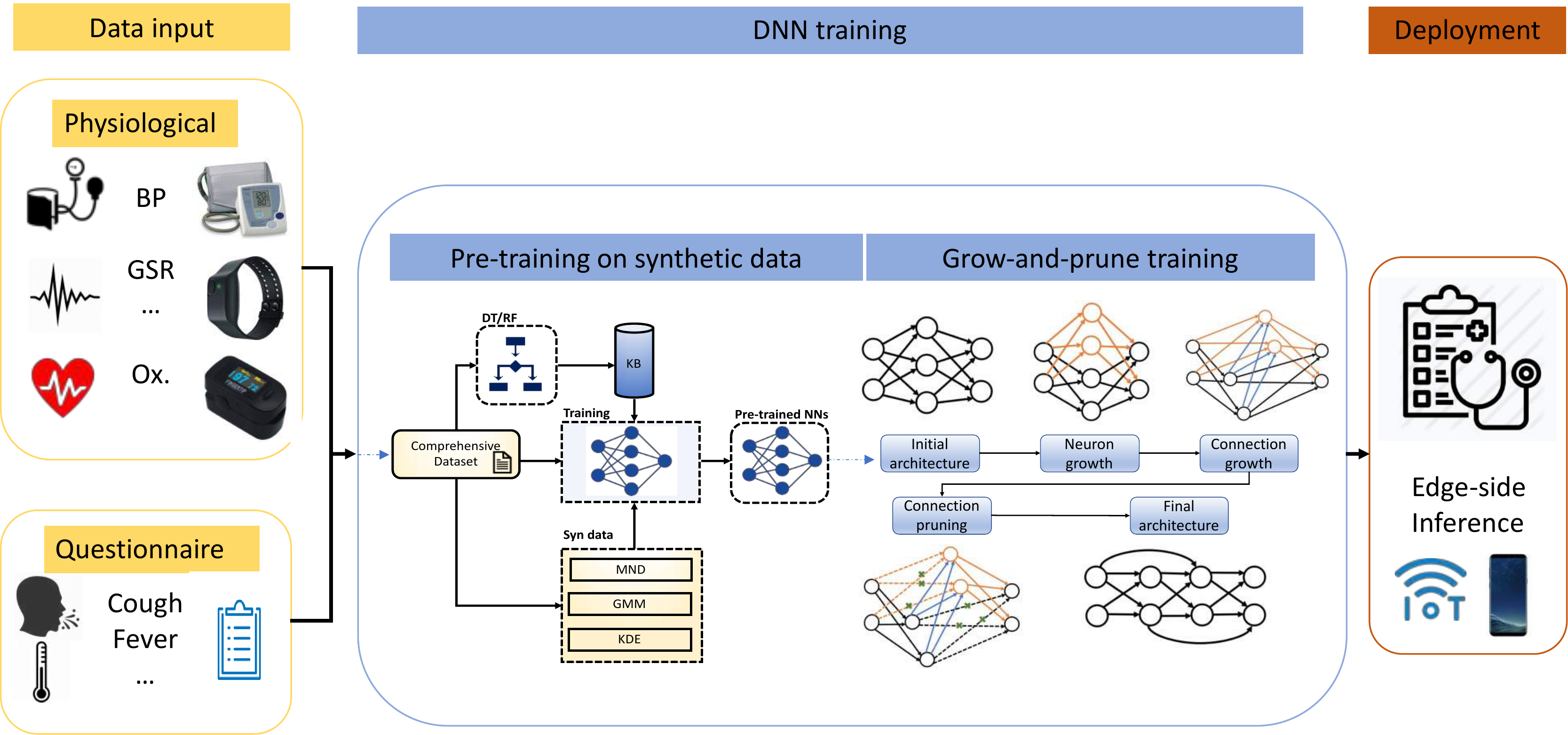}
    \caption{Schematic diagram of the CovidDeep framework (GSR: Galvanic skin response, 
IBI: inter-beat interval, Ox.: oxygen saturation, BP: blood pressure, DT/RF: decision tree/random 
forest, NN: neural network, KB: knowledge-base, MND: multi-variate Normal distribution, 
GMM: Gaussian mixture model, KDE: kernel density estimation).}
\label{fig:schematic}
\end{figure*}

\subsection{Framework overview}
The CovidDeep framework is shown in Fig.~\ref{fig:schematic}. CovidDeep obtains data from two 
different sources: physiological signals and questionnaire.  It has two flows: one that does
not use synthetic data and another one that does. When synthetic data are not used, the
framework just uses the real dataset divided into three categories: training, validation, and
test.  It trains the DNNs with the training dataset and picks the best one for the given set of
features based on the validation dataset, and finally tests this DNN on the test dataset to
obtain the test accuracy.  However, when the real training dataset size is small, it is often
advantageous to draw a synthetic dataset from the same probability distribution.
CovidDeep uses synthetic data generation methods to increase the dataset size and use
such data to pre-train the DNN architecture.  Then, it uses grow-and-prune synthesis to generate 
inference models that are both accurate and computationally-efficient. The models generated by 
CovidDeep are efficient enough to be deployed on the edge, e.g., the smartphone or smartwatch,
for SARS-CoV-2/COVID-19 inference.  

Next, we discuss the data input, model training, and model inference details.

\begin{itemize}
    \item \textbf{Data input}: As mentioned above, physiological signals and a questionnaire are
the two sources of data input to the model.  The physiological signals are derived from WMSs 
embedded in a smartwatch as well as a discrete pulse oximeter and blood pressure monitor. These 
signals can be easily obtained in a non-invasive, passive, and user-transparent manner. The list of 
these signals includes Galvanic skin response (GSR), inter-beat interval (IBI) that indicates
the heart rate, skin temperature, oxygen saturation, and blood pressure (systolic and diastolic).  
In the questionnaire, we asked the following yes/no questions: immune-compromised,
chronic lung disease, cough, shortness of breath, chills, fever, muscle pain, headache, sore
throat, smell-taste loss, and diarrhea.  We collected data on age, gender, weight, height, and 
smoking/drinking (yes/no), but did not find them to be useful either because of overfitting or
being unrepresentative.  All the relevant data sources are aggregated into a comprehensive data input 
for further processing.  
    \item \textbf{Model training}: CovidDeep uses different types of DNN models: (i) those
trained on the raw data only, (ii) those trained on raw data augmented with synthetic data to
boost accuracy, and (iii) those subjected to grow-and-prune synthesis for both boosting accuracy 
further and reducing model size.  The first type of DNN model uses a few hidden layers.  The
second type of DNN model is trained based on a system called TUTOR \cite{hassantabar2020Tutor} and 
is suitable for settings where data availability is limited.  It provides the DNN with a suitable 
inductive bias.  The third type of DNN model is based on the grow-and-prune DNN synthesis paradigm 
and employs three architecture-changing operations: neuron growth, connection growth, and connection 
pruning. These operations have been shown to yield DNNs that are both accurate and efficient 
\cite{hassantabar2019scann}. 
    \item \textbf{Model inference}: CovidDeep enables the users to have SARS-CoV-2/COVID-19 detection 
decision on their edge device on demand.
\end{itemize}

Next, we discuss the CovidDeep DNN architecture. 


\subsection{Model architecture}
\begin{figure}[!ht]
    \centering
    \includegraphics[scale= 0.5]{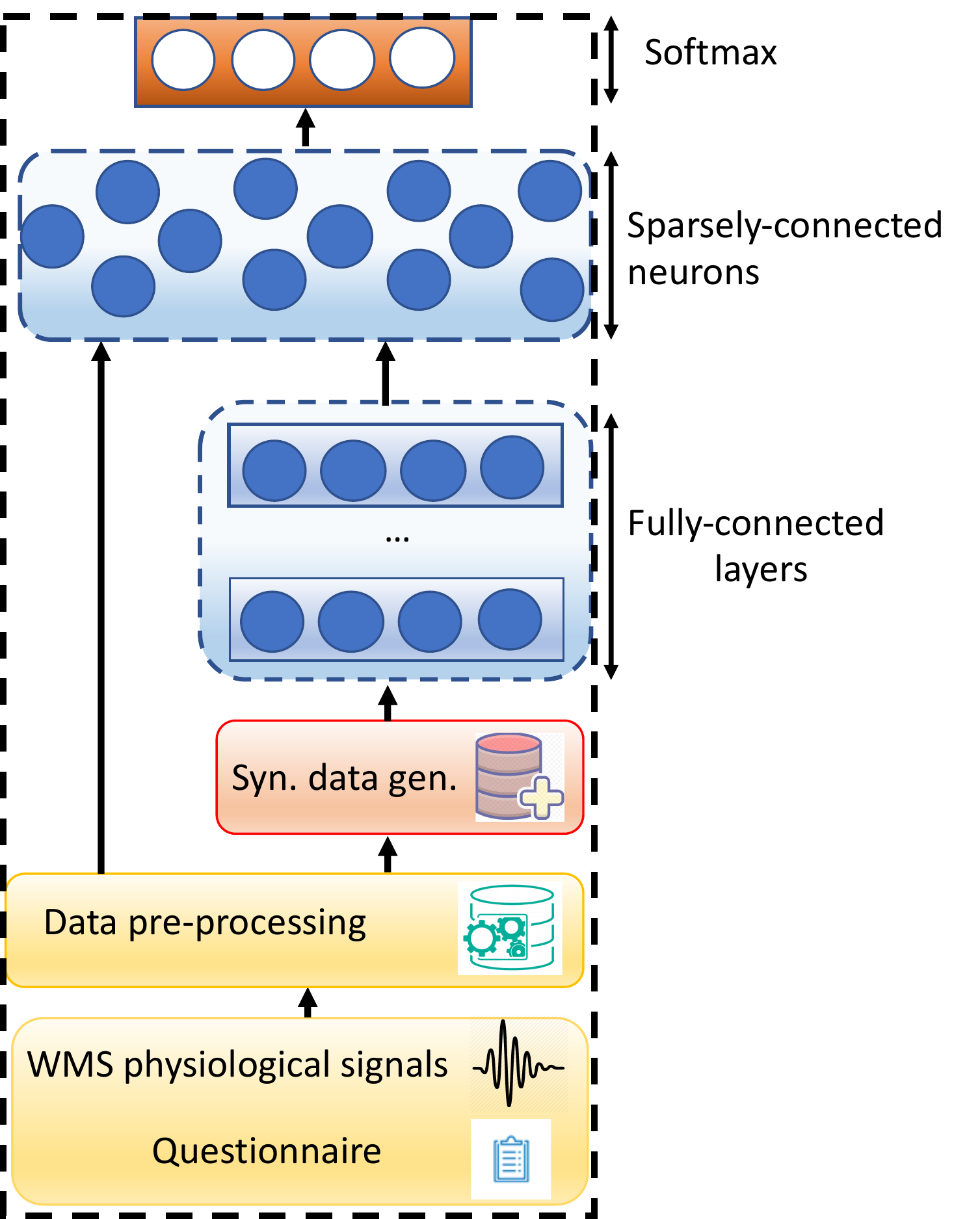}
    \caption{An illustration of the CovidDeep processing pipeline to generate predictions from data 
inputs.}
\label{fig:arch}
\end{figure}

Fig.~\ref{fig:arch} shows the processing pipeline of the CovidDeep framework. The architecture takes 
the data inputs (shown at the bottom) and generates a prediction, i.e., the detection decision, 
(shown at the top).  The pipeline consists of four steps: data pre-processing, synthetic data 
generation and architecture pre-training, grow-and-prune synthesis, and output generation 
through softmax. 

In the data pre-processing stage, data normalization and data alignment/aggregation are done.

\begin{itemize}
    \item \emph{Data normalization}: This step is aimed at changing feature values to a common
scale.  While data normalization is not always required, it is highly beneficial in the case of 
datasets that have features with very different ranges. It leads to better noise tolerance and 
improvement in model accuracy \cite{krizhevsky2012imagenet}. Data normalization can be done in 
several ways, such as min-max scaling and standardization. In this work, we use min-max scaling to 
map each data input to the $[0,1]$ interval. Scaling can be done as follows:
    \[
    x_{scaled} = \frac{x - \text{min($x$)}}{\text{max($x$) $-$ \text{min($x$)}}}
    \]
    
    \item \emph{Data alignment/aggregation}: The data from different WMSs may have different 
start times and frequencies. In order to merge them into a dataset, we need to synchronize the data 
streams based on their timestamps. The answers to the questions in the questionnaire are also added 
to the final dataset.
\end{itemize}

\noindent
\textbf{Synthetic data generation}: The training dataset generated in the above manner is next used to 
generate a synthetic dataset that is used to pre-train the DNN. These synthetic data and pre-training 
steps are based on the TUTOR framework \cite{hassantabar2020Tutor}.  The schematic diagram of 
the training scheme based on synthetic data is shown in Fig.~\ref{fig:syn-training}.  The synthetic 
dataset is generated in three different ways in TUTOR:

\begin{figure}[!ht]
    \centering
    \includegraphics[scale= 0.4]{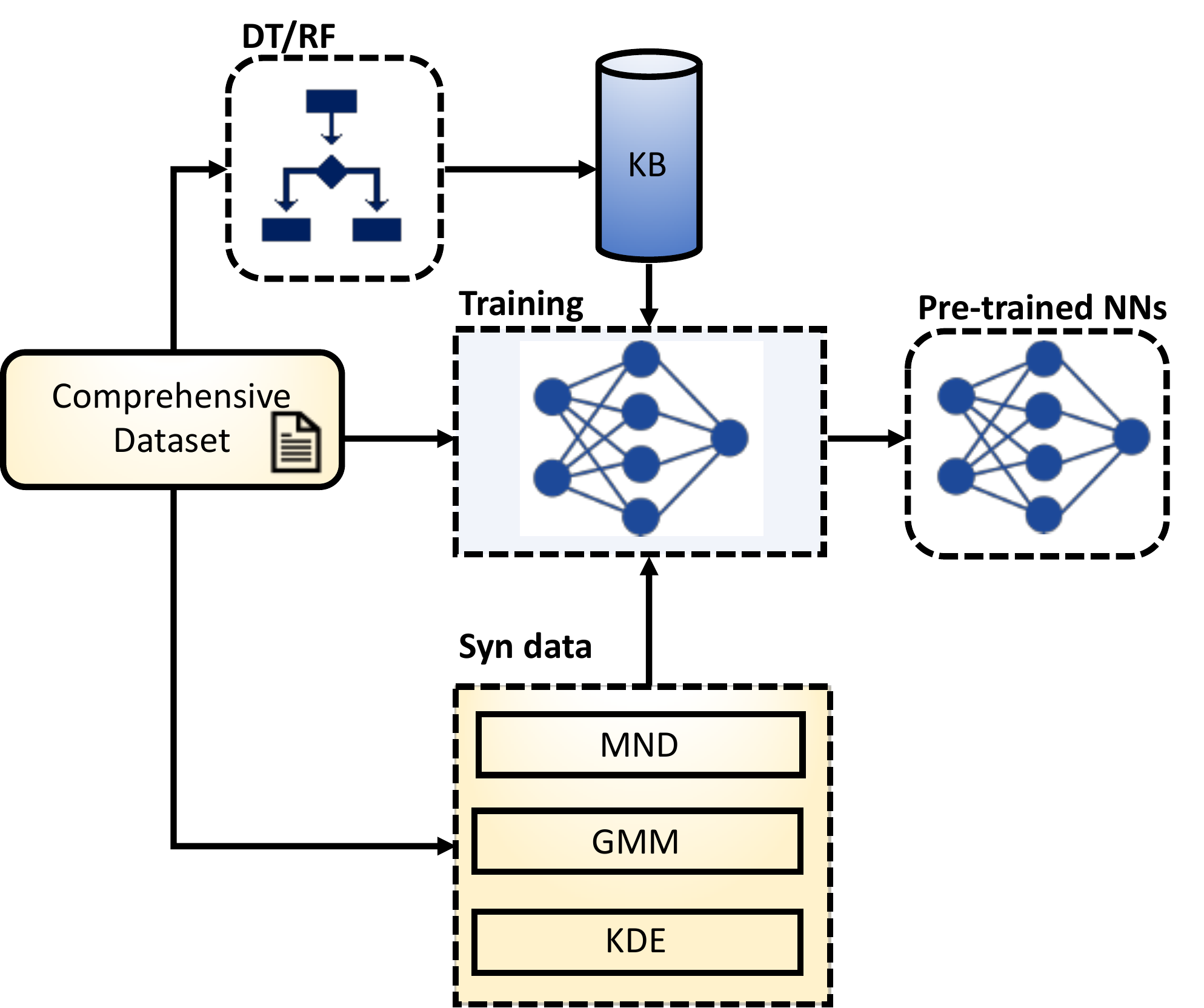}
    \caption{The schematic diagram for pre-training of the DNN model with the synthetic dataset 
(DT/RF: decision tree/random forest, NN: neural network, KB: knowledge-base).}
\label{fig:syn-training}
\end{figure}

\begin{itemize}
    \item Using multi-variate Normal distribution (MND): In this approach, the real training dataset,
i.e., the one obtained as a fraction of the data obtained from the WMSs and questionnaire, is modeled 
as a normal distribution to generate the synthetic data.
    \item Using Gaussian mixture model (GMM): This approach uses a multi-dimensional GMM to model the 
data distribution. The optimal number of GMM components is obtained with the help of a validation 
dataset. Subsequently, the synthetic dataset is generated from this GMM. 
    \item Using kernel density estimation (KDE): This approach uses non-parametric density estimation 
to estimate the probability distribution as a sum of many kernels. In our implementation, KDE is 
based on the Gaussian kernel function. The synthetic data are generated based on samples generated 
from this model.
\end{itemize}

\noindent
\textbf{Building a knowledge base (KB)}:
After generation of the synthetic data, we need to label the data points. To this end, we build a 
KB from the real training dataset. Decision tree (DT) and random forest (RF) are two 
classical machine learning methods that are inherently rule-based. In fact, each decision path in a 
decision tree, from the root to a leaf, can be thought of as a rule. Therefore, we aim to identify 
the set of rules that best describes the data.  We use such a model as a KB to label the generated 
synthetic dataset. 

\noindent
\textbf{Training with synthetic data}:
We use the labeled synthetic data to impose a prior on the DNN weights. To accomplish this, we 
pre-train the DNN model by using the generated synthetic dataset.  This provides the network
with an appropriate inductive bias and helps the network to ``get underway." 
This helps improve accuracy when data availability is limited.

\subsection{Grow-and-prune synthesis of the DNN}
In this section, we discuss the grow-and-prune synthesis paradigm 
\cite{dai2017nest,hassantabar2019scann}.  The approach presented in \cite{hassantabar2019scann} 
allows the depth of the DNN to grow during synthesis.  Thus, a hidden neuron can receive inputs from 
any neuron activated before it (including input neurons) and can feed its output to any neuron 
activated after it (including output neurons). As a result, the depth of the model is determined 
based on how the hidden neurons are connected, enabling the depth to be changed during 
training. We use three basic architecture-changing operations in the grow-and-prune synthesis process
that are discussed next. 

\noindent
\textbf{Connection growth}: 
This activates the dormant connections in the network. The weights of the added connections are set 
to $0$ and trained later.  We use two different methods for connection growth:
\begin{itemize}
    \item \textbf{Gradient-based growth}: This approach was first introduced by Dai et 
al.~\cite{dai2017nest}. Algorithm \ref{alg:gradient-growth} shows the process of gradient-based 
growth. Each weight matrix has a corresponding binary mask of the same size. This mask is used to 
disregard the inactive connections. The algorithm adds connections to reduce the loss function 
$\mathcal{L}$ significantly. To this end, the gradients of all the dormant connections are evaluated 
and their effectiveness ranked based on this metric. During a training epoch, the gradients of all 
the weight matrices for all the data mini-batches are captured in the back-propagation step. An 
inactive connection is activated if its gradient magnitude is large relative to the gradients in
its associated layer.
    \item \textbf{Full growth}: This connection growth restores all the dormant connections in
the network to make the DNN fully-connected.
\end{itemize}

\begin{algorithm}[h]
    \caption{Connection growth algorithm}
    \label{alg:gradient-growth}
    \begin{algorithmic}[l]
        \REQUIRE 
        $W \in R^{M \times N}$: weight matrix of dimension $M \times N$ (connecting layer with $M$ neurons to layer with $N$ neurons);
        $Mask \in R^{M \times N}$: weight mask of the same dimension as the weight matrix;
        Network $P$; $W.grad$: gradient of the weight matrix (of dimension $M \times N$); data $D$; 
        $\alpha$: growth ratio
        \IF{full growth}
        \STATE $Mask_{[1:M, 1:N]} = 1 $
        \ELSIF{gradient-based growth}
        \STATE {Forward propagation of data $D$ through network $P$ and then back propagation}
        \STATE {Accumulation of $W.grad$ for one training epoch}
        \STATE {$t = (\alpha \times MN)^{th}$ largest element in the $\left|W.grad\right|$ matrix} 
        
        \FORALL {$w.grad_{ij}$}
        \IF{$\left| w.grad_{ij} \right| > t$}
        \STATE {$Mask_{ij} = 1$}
        \ENDIF
        \ENDFOR
        \ENDIF
        \STATE $W$ = $W \otimes Mask$ 
        \ENSURE Modified weight matrix $W$ and mask matrix $Mask$
    \end{algorithmic}
\end{algorithm}

\noindent
\textbf{Connection pruning}: Connection pruning deactivates the connections that are smaller than a 
specified threshold. Algorithm \ref{alg:pruning} shows this process.

\begin{algorithm}[h]
    \caption{Connection pruning algorithm}
    \label{alg:pruning}
    \begin{algorithmic}[l]
        \REQUIRE Weight matrix $W \in R^{M \times N}$; mask matrix $Mask$ of the same dimension as 
        the weight matrix; $\alpha$: pruning ratio
        \STATE $t = (\alpha \times MN) ^{th}$ largest element in $\left|W\right|$
        \FORALL {$w_{ij}$}
        \IF{$\left| w_{ij} \right| < t$}
        \STATE {$Mask_{ij} = 0$}
        \ENDIF
        \ENDFOR
        \STATE $W$ = $W \otimes Mask$ 
        \ENSURE Modified weight matrix $W$ and mask matrix $Mask$
    \end{algorithmic}
\end{algorithm}

\noindent
\textbf{Neuron growth}: This step adds neurons to the network and thus increases network size.
This is done by duplicating existing neurons in the architecture.  To break the symmetry,
random noise is added to the weights of all the connections related to the newly added neurons. 
The neurons to be duplicated are either selected randomly or based on higher activation values. 
The process is explained in Algorithm \ref{alg:neuron-growth}.

\begin{algorithm}[h]
    \caption{Neuron growth algorithm}
    \label{alg:neuron-growth}
    \begin{algorithmic}[l]
        \REQUIRE Network $P$; weight matrix $W \in R^{M \times N}$; mask matrix $Mask$ of the same 
        dimension as the weight matrix; data $D$; candidate neuron $n_j$ to be added; array $A$ of activation values for all hidden neurons 
        \IF{activation-based selection}
        \STATE {forward propagation through $P$ using data $D$}
        \STATE {$i = argmax~(A)$}
        \ELSIF{random selection}
        \STATE {randomly pick an active neuron $n_i$}
        \ENDIF
        \STATE {$Mask_{j\cdot} = Mask_{i\cdot}, Mask_{{\cdot}j} = Mask_{{\cdot}i}$}
        \STATE {$w_{j\cdot} = w_{i\cdot} + noise, w_{{\cdot}j} = w_{{\cdot}i} + noise$}
        
        \ENSURE Modified weight matrix $W$ and mask matrix $Mask$
    \end{algorithmic}
\end{algorithm}

We apply connection pruning after neuron growth and connection growth in each iteration. 
Grow-and-prune synthesis starts from a fully connected architecture (mask values set to 1) and runs for a pre-defined number of iterations. 
Finally, the architecture
that performs the best on the validation dataset is chosen. 

\section{Implementation Details}
\label{sect:implementation}
In this section, we first explain how the data were obtained from 87 individuals and how various
datasets were prepared from the data.  We also provide implementation details of 
the CovidDeep DNN model.

\subsection{Data collection and preparation}
We collected physiological signals and questionnaire data 
with Institutional Research Board (IRB) approval at San Matteo Hospital in Pavia, Italy.
30 individuals were healthy (referred to as Cohort $1$) and the remaining were SARS-CoV-2-positive 
with varying levels of disease severity.  The SARS-CoV-2-positive cases were categorized into two
other cohorts: asymptomatic (Cohort $2$ with 27 individuals) and symptomatic
(Cohort $3$ with 30 individuals).  Distinguishing among
these cohorts is important to ascertain who may be spreading the virus unknowingly and to determine whether
medical support is needed for symptomatic individuals.  Hence, we 
train DNN models that can perform three-way classification.

To collect the physiological signals, we used commercially available devices: Empatica E$4$ 
smartwatch (sensors we found useful: GSR, IBI, skin temperature), a pulse oximeter, and a blood
pressure monitor.
Alongside the physiological signals, we employed a questionnaire to collect information about
possible COVID-$19$-related symptoms from all the individuals.  We also collected data
about age, gender, weight, height, and smoking/drinking (yes/no), but did not rely on these features 
as they were not necessarily representative of the larger population. 
Table \ref{tab:data} shows all the data types that we found to be useful.  The 
smartwatch data capture the physiological state of the user.  GSR measures continuous 
variations in the electrical characteristics of the skin, such as conductance, which can be caused by 
variations in body sweat.  IBI correlates with cardiac health. Furthermore, skin acts as a medium for 
insulation, sweat, and control of blood flow. Although it is not a clear indicator of internal body 
temperature, skin temperature helps assess skin health. 
The pulse oximeter indirectly measures blood oxygen saturation. 
It is a comfortable and painless way of measuring how well oxygen is being sent to parts of the 
body furthest from the heart, such as the arms and legs.  Blood pressure exposes various
underlying health problems.  Last, but not the least, the questionnaire elicits information that 
may help improve COVID-19 detection accuracy.  From all these sources of 
data, we derive various subsets as datasets for use in the CovidDeep framework to see 
which data features are the most beneficial to obtaining a high detection accuracy.  In
addition, the various sensor subsets have different costs. Hence, our results
also let one take test accuracy vs.~cost into consideration.

Before data collection commences, we inform the participants about the procedure. We then collect 
some relevant information and COVID-$19$-related symptoms in response to a questionnaire. 
We place the pulse oximeter on the index finger of the user for blood oxygen measurement. We also 
obtain the systolic/diastolic blood pressure measurements.  We place the smartwatch on the 
participant's wrist.  Data collection lasts for at most one hour for each participant, during 
which time we collect sensor data from the smartwatch.  We stream the data from the smartwatch to the 
smartphone over Bluetooth in real-time using a smartphone application. This application collects the 
data and performs basic validation to ensure data integrity. 

Next, we pre-process the raw data to generate a comprehensive dataset.  To this end, we first 
synchronize the WMS data streams.  We then divide the data streams into $15$-second data
windows. We then split the participants into three different sets of training, validation, and
test. The training set contains data from $52$ individuals, approximately $60\%$ of all the 
participants. Among the $52$ individuals represented in the training set, 18 are healthy, 16 are 
asymptomatic (but virus-positive), and 18 are symptomatic (and virus-positive). The validation set 
consists of data from 17 individuals, approximately $20\%$ of all the participants, with $6$, $5$, and 
$6$ individuals from Cohorts $1$, $2$, and 3, respectively. The test set contains data from 18 
individuals, approximately $20\%$ of all the participants, with $6$ individuals from each of the 
three cohorts.  This data partitioning ensures that all the data collected from any individual
are limited to just one of the three sets.  Furthermore, the data instances extracted from each 
individual have no time overlap.  In addition, in order to conduct ablation studies to gauge the 
impact of different data streams, we create different datasets, with various subsets of all the 
features. 

\begin{table}[]
\caption{Data types collected in the CovidDeep framework}
\label{tab:data}
\centering
\begin{tabular}{ll}
\toprule
Data type              & Data source   \\ 
\toprule
Immune-compromised     & Questionnaire \\
Chronic lung disease   & Questionnaire \\
Shortness of breath    & Questionnaire \\ 
Cough                  & Questionnaire \\
Fever                  & Questionnaire \\
Muscle pain            & Questionnaire \\
Chills                 & Questionnaire \\
Headache               & Questionnaire \\
Sore throat            & Questionnaire \\
Smell/taste loss       & Questionnaire \\
Diarrhea               & Questionnaire \\
\midrule
Galvanic skin response ($\mu$S)& Smartwatch   \\
Skin temperature ($^\circ C$) & Smartwatch   \\ 
Inter-beat interval ($ms$) & Smartwatch   \\ 
\midrule
Oxygen saturation (\%)& Pulse oximeter \\
Systolic blood pressure (mmHg) & Blood pressure monitor\\
Diastolic blood pressure (mmHg) & Blood pressure monitor\\
\bottomrule
\end{tabular}
\end{table}


\subsection{Model implementation}
We have implemented the CovidDeep framework in PyTorch. We perform DNN training on the Nvidia Tesla 
P$100$ data center accelerator, with $16$GB of memory.  We use cuDNN library to accelerate GPU
processing. Next, we give the details of the implemented DNN architectures trained on the
different datasets. 

We train various DNNs (with different numbers of layers and different numbers of
neurons per layer) and verify their performance on the validation dataset.
In general, a four-layer architecture with 256, 128, 128, and 3 neurons, respectively, performs the best.
The number of neurons in the input layer depends on which subset of features is selected for
training the DNN.  In the case of the full dataset, the input layer has 194 neurons, which
indicates the dataset dimension.  We obtain the features of the dataset from the 15$s$ 
data window as follows.  Sensor data collected from the smartwatch in the data window consist of 
180 signal readings, hence 180 features, from the three data streams running at $4$Hz.
We derive 11 features from the 11 questionnaire questions. Finally, we append the pulse oximeter 
oxygen saturation measurement and systolic/diastolic blood pressure measurements to obtain a 
feature vector of length 194.  

We use leaky ReLU as the nonlinear activation function in all the DNN layers.  As explained in 
Section \ref{sect:methodology}, we generate three DNNs for each dataset: (i) DNN
trained on the real training dataset, (ii) DNN pre-trained on the synthetic dataset
and then trained on the real training dataset, and (iii) DNN synthesized and trained with the
grow-and-prune synthesis paradigm.

\subsection{Network training}
\label{sect:training}
We use the stochastic gradient descent optimizer for DNN training, with a learning rate of $5$e-$3$ and batch size of $256$.  
We use $100000$ synthetic data 
instances to pre-train the network architecture.  Moreover, in the grow-and-prune synthesis
phase, we train the network for $20$ epochs each time the architecture changes.  We apply
network-changing operations over five iterations.  
In this step, we use pruning to achieve a pre-defined number of connections in the network, chosen 
based on performance on the validation set.  

\section{Experimental Results}
\label{sect:results}
In this section, we analyze the performance of CovidDeep DNN models.
We target three-way classification among the three cohorts described earlier.
In addition, we perform an ablation study to analyze the impact of different subsets of features 
as well as different steps of CovidDeep DNN synthesis. 

The CovidDeep DNN models are evaluated with four different metrics: test accuracy, false positive
rate (FPR), false negative rate (FNR), and F$1$ score. 
These terms are based on the following:
\begin{itemize}
    \item True positive (negative): SARS-CoV-2/COVID-$19$ (healthy) data instances classified as 
SARS-CoV-2/COVID-$19$ (healthy).
    \item False positive (negative): healthy (SARS-CoV-2/COVID-$19$) data instances classified as 
SARS-CoV-2/COVID-$19$ (healthy).
\end{itemize}
These metrics evaluate the model performance from different perspectives. Test accuracy
evaluates its overall prediction power.  It is simply the ratio of all the correct predictions
on the test data instances and the total number of such instances.
The FPR is defined as the ratio of the number of negative, 
i.e., healthy, instances wrongly categorized as positive (false positives) and the total number of 
actual negative instances.  The FNR is the ratio of positives that yield different test outcomes.  Thus, there is an FNR for both Cohorts 2 and 3.
Because of the three-way classification, the F$1$ score we report is the Macro F1 score.

\subsection{Model performance evaluation}
We obtained the highest test accuracy with a DNN model trained with the
grow-and-prune synthesis paradigm on the dataset that contained features from four
categories: GSR, pulse oximeter (Ox), blood pressure (BP), and questionnaire (Q).  
Table \ref{tab:confusion-GP} shows the confusion matrix for three-way classification 
among the three cohorts: Cohort 1 (healthy), Cohort 2 (asymptomatic-positive), Cohort 3
(symptomatic-positive), denoted as C1, C2, and C3, respectively. CovidDeep DNN achieves a test 
accuracy of 98.1\%.  The model achieves an FPR of only 0.8\%. The low FPR means that the model 
does not raise many false alarms.  It results in a 4.5\% FNR for Cohort 2 and a 0.0\% FNR for Cohort 
3, denoted as FNR(2) and FNR(3), respectively (each FNR refers to the ratio of the number of false 
predictions for that cohort divided by the total number of data instances of that type).  
The low FNRs demonstrate the ability of the DNN model to not miss virus-positive 
cases.  Moreover, the Macro F1 score of the DNN model is also high: 98.2\%.

\begin{table}[]
\caption{Confusion matrix for the most accurate three-way classification model}
\label{tab:confusion-GP}
\centering
\begin{tabular}{c|ccc|c}
\toprule
Label$\downarrow$\textbackslash Prediction$\rightarrow$ & C1 & C2 & C3 & Total\\
\toprule
C1 & $1066$ & $9$   & $0$    & $1075$ \\
C2 & $54$   & $1152$ & $0$    & $1206$ \\
C3 & $0$    & $0$    & $975$   & $975$ \\
\hline
Total& $1120$ & $1161$ & $975$ & $3256$ \\
\bottomrule
\end{tabular}
\end{table}

Next, we compare the three DNN models, trained on the real training dataset, with the aid of
synthetic data, and with the aid of grow-and-prune synthesis, for the most accurate case in 
Table \ref{tab:confusion-3DNNs}.  From this comparison, we see that the use of synthetic data and 
then grow-and-prune synthesis is able to boost the test accuracy compared to the DNN model trained on 
just the real dataset.  In addition, we see improvements in the FPR and FNR values.  The F1 score 
also follows the same trend, increasing with the use of synthetic data, and even more with the use of 
grow-and-prune synthesis. 

\begin{table*}[]
\caption{Test accuracy, FPR, FNRs, and F1 score (all in \%) for the three DNN models obtained for the 
most accurate case}
\label{tab:confusion-3DNNs}
\centering
\begin{tabular}{l|ccccc}
\toprule
DNN model trained on& Acc. & FPR & FNR(2) & FNR(3) & F1 Score\\
\toprule
Real training dataset & $79.9$ & $22.5$ & $34.2$ & $0.0$ & $80.9$\\
Real+synthetic training dataset & $84.8$ & $14.1$ & $28.4$  & $0.0$ & $85.5$ \\
Real+synthetic training dataset + grow-prune& $98.1$ & $0.8$ & $4.5$  & $0.0$ & $98.2$ \\
\bottomrule
\end{tabular}
\end{table*}


\subsection{Ablation studies}
In this section, we report results on various ablation studies. 
We begin by considering DNN models trained on features obtained from subsets of the six data
categories (five sensors and the questionnaire).  This helps us understand the impact of 
each of these categories and their various combinations.  
Then, we analyze the impact of different parts of the CovidDeep training process, pre-training 
with synthetic data, and grow-and-prune synthesis. 

Since there are six data categories from which the corresponding features are obtained, there
are 64 subsets.  However, one of these subsets is the null subset.  Thus, we evaluate the
remaining 63 subsets.  
For these evaluations, we only consider the first two types of DNN models, referred to as DNN
Models 1 and 2.  We consider grow-and-prune synthesis-based models later.
The results shown in Table~\ref{tab:feature-ablation} correspond to the case when features from
only one, two or three data categories are chosen, and in Table~\ref{tab:feature-ablation2} when 
features from four, five or six data categories are chosen.

We first notice that DNN Model 2 generally performs better than DNN Model 1 across the various
performance metrics.  This underscores the importance of using synthetic data when the available
dataset size is not large. Second, we observe that since this is a three-way classification,
only 33.3\% accuracy is possible by randomly predicting one of the three Cohorts.  Thus, even
single data categories (GSR, Temp, IBI, Ox, BP, Q) enable much better prediction than by chance.
These single data categories are still only weak learners of the correct label, 
when used in isolation.  
Third, DNN models, in general, tend to perform better on the 
various performance metrics when more data categories are used.  However, this is not always
true.  
For example, we obtain the highest accuracy of 93.6\% with DNN Model 2 when only
features from four (GSR, Temp, Ox, BP) of the six categories are used. Adding features based on IBI or 
Q or both to these four categories actually reduces the test accuracy.  
This may be due to the curse of dimensionality. 
When the number of features increases, in general, the dataset size needs to be increased to
obtain a good accuracy.  
For a fixed dataset size, this curse indicates that the number of features 
should be reduced.  
However, throwing out informative features would also reduce accuracy.  
In addition, some features are interactive, i.e., work synergistically to increase accuracy.  
Hence, a balance has to be found between accuracy and the number of features.  
Finally, when not all sensors are available (perhaps due to cost reasons), a suitable set that still provides reasonable
accuracy can be chosen based on the given cost budget.  
This may help a broader cross-section of the population access the technology.

\begin{table*}[]
\caption{Test accuracy, FPR, FNRs, and F1 score (all in \%) for two DNN models obtained for feature
subsets from one, two or three data categories}
\label{tab:feature-ablation}
\centering
\begin{tabular}{l|ccccc|ccccc}
\toprule
             & \multicolumn{5}{c|}{DNN Model 1} & \multicolumn{5}{c}{DNN Model 2}\\
Data category& Acc. & FPR & FNR(2) & FNR(3) & F1 Score & Acc. & FPR & FNR(2) & FNR(3) & F1 Score\\
\toprule
GSR  & $54.2$   & $22.1$   & $23.3$  & $99.6$ & $44.6$  & $54.2$ & $22.1$  & $23.4$ & $99.5$  & $44.7$ \\
Temp   & $57.2$   & $31.5$   & $60.3$  & $33.4$ & $57.5$  & $58.6$ & $32.2$  & $60.2$ & $28.2$  & $58.7$ \\
IBI   & $66.6$   & $55.1$   & $24.0$  & $21.1$ & $65.6$  & $66.8$ & $53.1$  & $25.1$ & $21.1$  & $66.0$ \\
Ox   & $45.4$   & $56.2$   & $59.6$  & $46.7$ & $45.5$  & $45.4$ & $56.2$  & $59.6$ & $46.7$  & $45.5$ \\
BP   & $44.3$   & $96.3$   & $60.3$  & $5.2$ & $36.4$  & $44.3$ & $96.3$  & $60.3$ & $5.2$  & $36.4$ \\
Q   & $61.4$   & $0.0$   & $100.0$  & $5.2$ & $53.5$  & $63.0$ & $0.0$  & $100.0$ & $0.0$  & $54.7$ \\
GSR+Temp   & $57.2$   & $33.4$   & $60.3$  & $31.4$ & $57.3$  & $76.9$ & $6.4$  & $44.1$ & $15.4$  & $76.5$ \\
GSR+IBI   & $74.9$ & $3.2$  & $34.6$ & $37.4$  & $74.3$ & $76.1$   & $3.6$   & $31.9$  & $36.3$ & $75.5$  \\
GSR+Ox   & $52.7$   & $29.0$   & $44.2$  & $71.3$ & $51.3$  & $47.5$ & $44.3$  & $44.7$ & $71.3$  & $46.1$ \\
GSR+BP   & $55.2$   & $70.7$   & $53.8$  & $5.2$ & $52.7$  & $64.1$ & $46.4$  & $51.2$ & $5.2$  & $63.7$ \\
GSR+Q   & $89.1$   & $6.8$   & $23.3$  & $0.0$ & $89.6$  & $89.2$   & $6.7$   & $23.3$  & $0.0$ & $89.7$ \\
Temp+IBI   & $68.1$   & $19.3$   & $53.9$  & $18.8$ & $68.4$  & $68.2$ & $19.9$  & $52.9$ & $18.9$  & $68.6$ \\
Temp+Ox   & $48.3$   & $26.3$   & $78.4$  & $46.7$ & $46.5$  & $49.3$ & $24.2$  & $77.7$ & $46.7$  & $47.3$ \\
Temp+BP   & $50.3$   & $84.5$   & $54.7$  & $5.2$ & $45.9$  & $53.7$ & $74.0$  & $54.7$ & $5.2$  & $50.9$ \\
Temp+Q   & $68.9$   & $26.5$   & $60.4$  & $0.0$ & $69.8$  & $69.0$ & $26.3$  & $60.3$ & $0.0$  & $69.9$ \\
IBI+Ox   & $48.1$   & $60.4$   & $68.0$  & $22.7$ & $49.8$  & $49.0$ & $58.3$  & $68.0$ & $22.1$  & $50.7$ \\
IBI+BP   & $47.8$   & $92.8$   & $54.0$  & $5.2$ & $44.8$  & $48.5$ & $89.8$  & $54.9$ & $5.2$  & $46.3$ \\
IBI+Q   & $80.9$   & $19.5$   & $34.2$  & $0.0$ & $81.8$  & $80.9$ & $17.8$  & $35.8$ & $0.0$  & $81.7$ \\
Ox+BP   & $59.6$   & $56.2$   & $54.8$  & $5.2$ & $59.1$  & $66.9$ & $56.2$  & $35.0$ & $5.2$  & $66.8$ \\
Ox+Q   & $50.2$   & $56.2$   & $80.2$  & $5.2$ & $52.5$  & $50.2$ & $56.2$  & $80.2$ & $5.2$  & $52.5$ \\
BP+Q   & $51.8$   & $56.2$   & $80.1$  & $0.0$ & $49.9$  &  $57.6$   & $56.2$   & $60.3$  & $5.2$ & $56.8$ \\
GSR+Temp+IBI   & $70.5$   & $11.5$   & $54.7$  & $17.9$ & $70.8$  & $76.6$ & $3.5$  & $46.0$ & $17.2$  & $76.7$ \\
GSR+Temp+Ox   & $69.1$   & $22.1$   & $33.5$  & $37.2$ & $70.0$  & $69.7$ & $23.1$  & $27.1$ & $42.4$  & $70.2$ \\
GSR+Temp+BP   & $57.0$   & $64.0$   & $54.8$  & $5.2$ & $55.4$  & $67.0$ & $34.2$  & $54.4$ & $5.2$  & $66.4$ \\
GSR+Temp+Q   & $83.6$   & $0.2$   & $44.2$  & $0.0$ & $83.9$  & $91.3$ & $0.2$  & $23.3$ & $0.0$  & $91.7$ \\
GSR+IBI+Ox   & $64.8$   & $14.0$   & $45.4$  & $45.8$ & $64.8$  & $70.8$ & $19.1$  & $43.2$ & $23.0$  & $71.7$ \\
GSR+IBI+BP   & $60.2$ & $34.4$  & $52.8$ & $29.5$  & $61.5$ & $64.3$   & $32.2$   & $43.7$  & $29.5$ & $64.8$  \\
GSR+IBI+Q   & $87.7$   & $11.2$   & $23.3$  & $0.0$ & $88.3$  & $88.8$   & $7.7$   & $23.3$  & $0.0$ & $89.4$ \\
GSR+Ox+BP   & $71.3$   & $40.7$   & $37.1$  & $5.2$ & $71.2$  & $81.9$ & $23.1$  & $4.1$ & $29.8$  & $82.1$ \\
GSR+Ox+Q   & $69.9$ & $22.9$  & $56.7$ & $5.2$  & $71.0$  & $75.5$   & $22.7$   & $41.8$  & $5.2$ & $76.7$ \\
GSR+BP+Q   & $63.9$   & $26.5$   & $73.8$  & $0.0$ & $62.3$  & $64.1$ & $25.9$  & $73.8$ & $0.0$  & $62.4$ \\
Temp+IBI+Ox   & $57.4$   & $38.9$   & $62.4$  & $22.2$ & $57.5$  & $61.8$ & $30.7$  & $57.8$ & $22.2$  & $61.8$ \\
Temp+IBI+BP   & $55.8$   & $71.6$   & $51.2$  & $5.2$ & $53.9$  & $55.3$ & $70.0$  & $54.0$ & $5.2$  & $53.0$ \\
Temp+IBI+Q   & $73.6$ & $17.2$  & $51.8$ & $5.0$  & $74.5$ & $77.1$   & $9.0$   & $53.6$  & $0.0$ & $77.5$ \\
Temp+Ox+BP   & $70.6$   & $34.5$   & $44.2$  & $5.4$ & $72.1$  & $72.3$ & $33.9$  & $40.4$ & $5.2$  & $73.7$ \\
Temp+Ox+Q   & $53.3$   & $56.2$   & $71.8$  & $5.2$ & $55.8$  & $53.4$ & $56.2$  & $71.4$ & $5.2$  & $55.9$ \\
Temp+BP+Q   & $47.9$   & $46.6$   & $94.9$  & $5.2$ & $43.5$  & $49.9$ & $40.8$  & $94.7$ & $5.2$  & $45.1$ \\
IBI+Ox+BP   & $65.0$   & $59.1$   & $37.5$  & $5.2$ & $66.1$  & $64.1$ & $60.8$  & $38.4$ & $5.2$  & $65.0$ \\
IBI+Ox+Q   & $54.8$   & $56.2$   & $67.8$  & $5.2$ & $57.2$  & $55.0$ & $56.2$  & $67.2$ & $5.2$  & $57.4$ \\
IBI+BP+Q   & $55.9$   & $56.2$   & $65.2$  & $4.6$ & $55.0$  & $53.4$ & $56.2$  & $71.6$ & $5.2$  & $52.3$ \\
Ox+BP+Q   & $66.9$   & $56.2$   & $35.0$  & $5.2$ & $68.2$  & $66.9$ & $56.2$  & $35.0$ & $5.2$  & $68.2$ \\
\bottomrule
\end{tabular}
\end{table*}

\begin{table*}[]
\caption{Test accuracy, FPR, FNRs, and F1 score (all in \%) for two DNN models obtained for feature 
subsets from four, five or six data categories}
\label{tab:feature-ablation2}
\centering
\begin{tabular}{l|ccccc|ccccc}
\toprule
             & \multicolumn{5}{c|}{DNN Model 1} & \multicolumn{5}{c}{DNN Model 2}\\
Data category& Acc. & FPR & FNR(2) & FNR(3) & F1 Score & Acc. & FPR & FNR(2) & FNR(3) & F1 Score\\
\toprule
GSR+Temp+IBI+Ox   & $76.6$   & $23.3$   & $27.0$  & $19.2$ & $77.3$  & $74.5$ & $28.5$  & $28.3$ & $18.8$  & $75.2$ \\
GSR+Temp+IBI+BP   & $62.5$   & $27.1$   & $53.4$  & $29.2$ & $62.4$  & $73.3$ & $13.6$  & $44.0$ & $19.8$  & $73.4$ \\
GSR+Temp+IBI+Q   & $87.1$   & $0.2$   & $34.7$  & $0.0$ & $87.5$  & $89.1$ & $1.6$  & $27.9$ & $0.0$  & $89.6$ \\
GSR+Temp+Ox+BP   & $77.6$   & $24.2$   & $34.7$  & $5.2$ & $77.8$  & $93.6$ & $1.7$  & $11.4$ & $5.2$  & $93.7$ \\
GSR+Temp+Ox+Q   & $80.7$ & $22.5$  & $27.8$ & $5.2$  & $81.7$ & $81.2$   & $22.5$   & $26.4$  & $5.2$ & $82.2$ \\
GSR+Temp+BP+Q   & $60.0$   & $11.5$   & $93.4$  & $5.2$ & $53.2$  & $61.8$ & $11.5$  & $93.0$ & $0.0$  & $54.5$ \\
GSR+IBI+Ox+BP   & $75.0$   & $23.3$   & $42.6$  & $5.2$ & $76.1$  & $76.8$ & $24.2$  & $37.0$ & $5.2$  & $77.8$ \\
GSR+IBI+Ox+Q   & $69.8$   & $32.2$   & $48.5$  & $5.2$ & $71.4$  & $76.1$ & $40.4$  & $24.5$ & $4.9$  & $77.1$ \\
GSR+IBI+BP+Q   & $59.3$   & $32.6$   & $80.3$  & $0.8$ & $57.1$  & $66.2$ & $3.4$  & $84.5$ & $4.6$  & $60.7$ \\
GSR+Ox+BP+Q   & $79.9$   & $22.5$   & $34.2$  & $0.0$ & $80.9$  & $84.8$ & $14.1$  & $28.4$ & $0.0$  & $85.5$ \\
Temp+IBI+Ox+BP   & $59.2$ & $52.9$  & $58.9$ & $5.2$  & $61.1$ & $66.9$   & $53.8$   & $37.2$  & $5.2$ & $67.9$   \\
Temp+IBI+Ox+Q   & $63.1$   & $48.5$   & $52.2$  & $5.2$ & $65.1$  & $62.1$ & $56.2$  & $48.0$ & $5.2$  & $64.0$ \\
Temp+IBI+BP+Q   & $54.5$   & $31.9$   & $90.3$  & $5.2$ & $49.8$  & $54.7$ & $30.7$  & $90.7$ & $5.1$  & $49.8$ \\
Temp+Ox+BP+Q   & $67.1$   & $56.2$   & $34.5$  & $5.2$ & $68.3$  & $66.8$ & $56.2$  & $35.3$ & $5.2$  & $68.1$ \\
IBI+Ox+BP+Q   & $66.9$   & $56.2$   & $35.0$  & $5.2$ & $68.2$  & $66.9$ & $56.2$  & $35.0$ & $5.2$  & $68.2$ \\
GSR+Temp+IBI+Ox+BP   & $77.1$   & $29.1$   & $31.8$  & $5.2$ & $78.2$  & $83.3$ & $34.2$  & $10.3$ & $5.2$  & $83.7$ \\
GSR+Temp+IBI+Ox+Q   & $67.2$   & $5.8$   & $79.1$  & $5.2$ & $65.3$  &  $83.1$ & $20.1$  & $23.5$ & $5.2$  & $83.9$ \\
GSR+Temp+IBI+BP+Q   & $64.3$   & $4.7$   & $88.2$  & $5.1$ & $57.8$  & $69.0$ & $15.7$  & $65.8$ & $4.7$  & $67.0$ \\
GSR+Temp+Ox+BP+Q   & $83.8$   & $0.4$   & $39.1$  & $5.2$ & $84.2$  & $83.8$ & $0.4$  & $39.1$ & $5.2$  & $84.2$ \\
GSR+IBI+Ox+BP+Q   & $71.8$   & $37.5$   & $38.5$  & $5.2$ & $73.3$  & $75.3$ & $23.8$  & $41.1$ & $5.2$  & $76.6$ \\
Temp+IBI+Ox+BP+Q    & $62.5$ & $44.8$  & $57.0$ & $5.2$  & $64.5$  & $66.6$   & $48.8$   & $42.4$  & $5.2$ & $68.3$ \\
GSR+Temp+IBI+Ox+BP+Q   & $77.8$   & $18.3$   & $39.4$  & $5.2$ & $78.8$  & $83.7$ & $26.9$  & $15.9$ & $5.2$  & $84.1$ \\
\bottomrule
\end{tabular}
\end{table*}

To illustrate the effect of the different parts of the CovidDeep training process, we compare
11 CovidDeep DNN models, trained based on the different DNN synthesis and training steps.   
We chose these models from different accuracy ranges.
Table~\ref{tab:NN-ablation3} shows comparison results for the three-way classification task. 
We have already compared various performance metrics for DNN Models 1 and 2 earlier.  Hence,
here, we just report their accuracy, FLOPs, and number of model parameters (\#Param). The best 
DNN Model 3 was obtained with the help of the validation dataset.  This enabled us to find the
best \#Param. value. Only this model was tested on the test dataset.
Acc.(1) and Acc.(2), respectively, refer to the accuracy of DNN Models 1 and 2. The FLOPs and
\#Param. for these two models are identical.  We report
all the performance metrics for DNN Model 3 that is generated by grow-and-prune synthesis using
both real and synthetic data.  Thus, the starting point for DNN Model 3 synthesis is DNN Model 2.
Next, we compare DNN Model 3 with the other two models based on various measures and show why
it is suitable for deployment on the edge devices.

\begin{itemize}
    \item \textbf{Smaller model size}: It contains $3.4\times$ fewer parameters on an average 
(geometric mean) than DNN Models 1 and 2, thus significantly reducing the memory requirements.
    \item \textbf{Less computation}: It reduces FLOPs per inference by $3.5\times$ on an average 
(geometric mean) relative to DNN Models 1 and 2, thus facilitating more efficient inference on the 
edge devices.
    \item \textbf{Better performance}: It improves accuracy on an average by $7.8$\% ($1.9$\%) 
relative to DNN Model 1 (2), while also lowering FPR and FNRs, in general.
\end{itemize}

\small\addtolength{\tabcolsep}{-1.5pt}
\begin{table*}[]
\caption{Comparison of the three DNN models (all performance metrics in \%) for various feature sets}
\label{tab:NN-ablation3}
\centering
\begin{tabular}{l|cccc|cccccccc}
\toprule
             & \multicolumn{4}{c|}{DNN Models 1 and 2} & \multicolumn{7}{c}{DNN Model 3}\\
Data category& Acc.(1) & Acc.(2) &FLOPs &\#Param. & Acc. &FLOPs &\#Param & FPR & FNR(2) & FNR(3) & F1 Score \\
\toprule
GSR+Ox+BP+Q  & 79.9 & 84.8 & 136.4k & 68.5k & 98.1 & 19.5k & 10.0k & 0.8 & 4.5 & 0.0 & 98.2 \\
GSR+IBI+Q        & 87.7 & 88.8 & 165.6k & 83.1k & 91.5 & 39.5k & 20.0k & 1.3 & 21.9 & 0.0 & 91.9 \\
GSR+Q    & 89.1 & 89.2 & 134.9k & 67.7k & 91.3 & 9.5k & 5.0k & 0.2 & 23.2 & 0.0 & 91.7 \\
GSR+Temp+Q & 83.6 & 91.3 & 165.6k & 83.1k & 91.3 & 151.5k & 76.0k & 0.2 & 23.3 & 0.0 & 91.7 \\
GSR+Temp+IBI+Q       & 87.1 & 89.1 & 196.3k & 98.4k & 90.7 &  19.5k & 10.0k & 0.2 & 20.7 & 5.2 & 91.0  \\
GSR+Temp+Ox+Q    & 80.7 & 81.2 & 166.1k & 83.3k & 87.7 & 119.5k & 60.0k & 0.3 & 28.7 & 5.2 & 88.1 \\
GSR-Temp-IBI-Ox-Q & 67.2 & 83.1 & 196.8k & 98.7k & 86.4 & 59.5k & 30.0k & 11.3 & 22.6 & 5.2 & 87.0 \\
GSR+Temp+IBI+Ox+BP & 77.1 & 83.3 & 192.2k & 96.4k & 84.6 & 59.5k & 30.0k & 29.5 & 11.2 & 5.2 & 85.1 \\
GSR+Ox+BP   & 71.3 & 81.9 & 130.8k & 65.7k &  82.4 & 89.5k & 45.0k & 23.8 & 2.1  & 29.8 & 82.5  \\
GSR+Temp+Ox+BP  & 77.6 & 93.6 & 161.5k & 81.0k & 82.3 & 129.5k & 65.0k & 25.2 & 21.0 & 5.2 & 82.8 \\
IBI+Q  & 80.9 & 80.9 & 134.9k & 67.7k & 81.7 & 19.5k & 10.0k & 29.3 & 23.3 & 0.0 & 82.5 \\
\bottomrule
\end{tabular}
\end{table*}

\section{Discussion and Future Work}
\label{discussion}

In this section, we discuss the inspirations we took from the human brain in the synthesis process of 
CovidDeep DNNs. We also discuss future directions in medical research enabled by the CovidDeep 
framework. 

An interesting ability of the human brain is to efficiently solve novel problems in a new domain 
despite limited prior experience.  Inspired by this human capability, CovidDeep 
uses the TUTOR \cite{hassantabar2020Tutor} approach for synthetic data generation and labeling to 
help the neural network start from a better initialization point.  Use of gradient descent from
a learned initialization point provides the DNN with an appropriate inductive bias.  Hence, it 
reduces the need for large datasets that are not readily available for SARS-CoV-2/COVID-$19$ 
AI research. 

The CovidDeep DNN training process takes another inspiration from the human brain development process 
in the grow-and-prune synthesis step.  The human brain undergoes dynamic changes in its synaptic 
connections every second of its lifetime. Acquisition of knowledge depends on these synaptic 
rewirings \cite{grossberg1988nonlinear}.  Inspired by this phenomenon, CovidDeep utilizes 
the grow-and-prune synthesis paradigm to enable DNN architecture adaptation throughout training.
CovidDeep DNNs synthesized with grow-and-prune synthesis do not suffer from the situation faced by 
most current DNNs: fixed connections during training.  This enables CovidDeep to generate 
very compact, yet accurate, models for SARS-CoV-2/COVID-$19$ detection. 

CovidDeep uses physiological signals extracted using commercially available devices and achieves 
high test accuracy. As a result, it provides a testing mechanism that is accurate, easily 
accessible to the general public, and easy for individuals to use.  Furthermore, this 
mechanism only requires a few minutes of data collection from an individual to perform an
inference. Note that at most one hour of data collection from each individual was only required for
training of the DNN models.  It does not require the presence of a nurse or physician during testing.  
In fact, besides the data collected by the smartwatch and discrete sensors (for obtaining
blood oxygen and blood pressure), the additional information required by the electronic questionnaire 
is small, related to the general health of the subject, and can be easily filled out with a yes/no 
answer.  Thus, CovidDeep has the potential to significantly decrease the spread of SARS-CoV-2, 
save hundreds of thousands of lives, and drastically reduce the need for hospitalization, 
while also helping the world economy recover.

CovidDeep demonstrates that WMS-based SARS-CoV-2/COVID-19 detection is feasible.  Previously, diabetes
diagnosis was shown to be possible with the help of such sensors \cite{yin2019diabdeep}. 
We believe that WMS-based disease detection is feasible for a large number of diseases
\cite{yin2017health}.

Since data were collected from only 87 individuals, despite being augmented with synthetic
training data drawn from the real training data probability distribution, more work is needed
for validating the various DNN models in the field, especially since the data were obtained
from a single location in Italy.  This process has begun across various continents.


\section{Conclusion}
\label{conclusion}
In this article, we proposed a framework called CovidDeep to facilitate daily and pervasive 
detection of SARS-CoV-2/COVID-19. The framework combines off-the-shelf WMSs with efficient DNNs to
achieve this goal.  CovidDeep DNNs can be easily deployed on edge devices (e.g., smartphones and 
smartwatches) as well as servers. CovidDeep uses synthetic data generation to alleviate the need for 
large datasets. In addition, training of CovidDeep DNNs based on the grow-and-prune synthesis 
paradigm enables them to learn both the weights and the architecture during training. 
CovidDeep was evaluated based on data collected from 87 individuals. The highest accuracy it
achieves is 98.1\%.  However, several subsets of features that correspond to easily accessible
sensors in the market also achieve high enough accuracy to be practically useful.  
With more data collected from larger deployment scenarios, the accuracy of CovidDeep DNNs can
be improved further through incremental learning.

\noindent
{\bf Contributions:}  The SARS-CoV-2/COVID-19 detection project was conceived by Niraj K. Jha.  He 
also supervised the dataset preparation and DNN model generation efforts. Shayan Hassantabar 
performed DNN synthesis and evaluation.  Vishweshwar Ghanakota developed the smartphone application 
for data collection, authenticated the credentials of the application sending data, ensured data
integrity, and ran pre-processing scripts.  Gregory N. Nicola MD and Ignazio R. Marino MD defined the 
patient cohorts, and helped with the IRB approval process. Gregory N. Nicola MD, Ignazio R. Marino 
MD, and Bruno Raffaele decided on the questions to be placed in the questionnaire.  Novati Stefano, 
Alessandra Ferrari, and Bruno Raffaele collected data from patients and healthy individuals and 
labeled the data.  Kenza Hamidouche helped with the synthesis and evaluation of the DNN models.
All co-authors helped with the revision and editing of the manuscript.

\noindent
{\bf Acknowledgments:} The project was facilitated by the tireless efforts of Bob Schena (CEO,
Rajant Corp.) and Adel Laoui (CEO, NeuTigers, Inc.). Giana Schena and Maria Schena helped with
buying and transporting the instruments as well as English-to-Italian translations of various 
documents.  Joe Zhang helped initially with feature extraction from the raw dataset.
Claudia Cirillo coordinated the administrative work and helped with translation of
documents to Italian for the IRB application. Ravi Jha helped with proofreading of the manuscript.The Chief of the Italian Police, Franco Gabrielli, helped ensure safe and fast entrance and transfer of US researchers on Italian soil during the
COVID-19 lockdown. 

\noindent
{\bf Competing interests:} 
Five of the co-authors of this article, Niraj K. Jha, Shayan Hassantabar, Vishweshwar 
Ghanakota, Gregory N. Nicola MD, and Kenza Hamidouche have equity in NeuTigers, Inc. Neutigers, along 
with Rajant Corporation and Thomas Jefferson University and Jefferson Health, enabled data collection 
from San Matteo Hospital, Pavia, Italy.

\bibliographystyle{IEEEtran}
\bibliography{CovidDeep}

\end{document}